\documentclass[prd,aps,twocolumn,showpacs,superscriptaddress]{revtex4-2}

\usepackage{amsmath,amssymb}
\usepackage{verbatim}
\usepackage{graphicx}
\usepackage{rotating}
\usepackage[pdftex,
      colorlinks=true,
      linkcolor=blue,
      urlcolor=cyan,
      filecolor=magenta,
      citecolor=blue,
      pdfstartview=FitV,
      pdftitle={},
        pdfauthor={Lalremruati, Lalrinfela, Zodinmawia},
        pdfsubject={},
        pdfkeywords={},
        pdfpagemode=None,
        bookmarksopen=true
      ]{hyperref}
\usepackage{caption}
\usepackage{subcaption}

\usepackage{graphicx}
\usepackage{dcolumn}
\usepackage{bm}

\begin{document}

\title{Tidal love number and its influence on the pericenter shift of S-stars near Sgr A* }

\author{P.C Lalremruati}
\email{pcremruati905@gmail.com}
\affiliation{Department of Physical Sciences, Indian Institute of Science Education and Research Kolkata, Mohanpur, Nadia - 741246, West Bengal, India}
\affiliation{Center of Excellence in Space Sciences India, Mohanpur, Nadia-741246, West Bengal, India}
\author{H Lalrinfela}
\affiliation{
 Department of Physics, Mizoram University, Aizawl, Tanhril -796004, Mizoram, India
}%
\author{Zodinmawia}
\affiliation{
 Department of Physics, Mizoram University, Aizawl, Tanhril -796004, Mizoram, India
}%

\date{\today}

\preprint{}

\begin{abstract} 
The tidal love number determines a star's deformability rate in the presence of gravitational potential and depends on the star's internal structure. In this work, we investigate two significant prospects on tidal love number : (i) the influence of the polytropic index of stars on the tidal love number and , (ii) how tidal love number affects the pericenter shift of S-stars near Sgr A* which is an important probe for strong-field tests of gravitational theories. We consider S-stars orbiting Sgr A* at a pericenter distance of 45 au to 500 au, well below the S-2 orbit. The S-stars have polytropic indices of the range, n = 1.0, 1.5, 2.0, 2.5, 3.0, 3.5, 4.0 and 4.5  and eccentricity, e = 0.9 inclined at $i=90 ^{\circ}$. The tidal love number is estimated for multipole moments $l=2$ and $l=3$. It has been found that the tidal love number decreases as the polytropic index increases. Additionally, the tidal love number for the multipole moment of $l=2$ is dominant over that of $l=3$. The tidal distortion effect also causes a greater pericenter shift in compact orbit S-stars with lower polytropic indices and tidal love number having multipole moment $l=2$. The estimated results offer relevant insights for testing general relativity and its alternative theories in the vicinity of Sgr A*.

 \end{abstract}


\maketitle

\section{ Introduction}
The supermassive black hole located at the center of our Milky Way galaxy, sitting at a distance of approximately 8kpc (equivalent to 26,000lyrs) \cite{Amorim2019} from us, provides an excellent opportunity to test several gravitational theories in extreme regions of gravity. This is due to its proximity, as well as the fact that it serves as a perfect platform for studying black holes and their effects on surrounding matter. The studies conducted by Genzel et al. in 1997 and Gillessen et al. in 2017 have shed more light on this fascinating phenomenon \cite{Genzelet1997, Gillessen2017, SalcidoRoldan2020}. Recent observations made by the GRAVITY Collaboration have estimated the mass of the  Galactic Center supermassive black hole, Sgr A* to be of the order  $\sim$ $4.25 \times 10^6$ $M_{\odot}$\cite{GRAVITY2020}.  Surrounding Sgr A* is a group of main-sequence B-type stars commonly referred to as S-stars \cite{Ghez1998, Genzel2003, Eisenhauer2003, Zakharov2018}. Over a decade, two teams of astronomers led by Andrea Ghez and Reinhard Genzel have closely monitored and studied the motion of S-stars \cite{Genzelet1997, Ghez1998, Zakharov2018}. Andrea Ghez and her team utilized the Keck telescope to study the S-stars, while Reinhard’s team used the GRAVITY detector in the Very Large Telescope (VLT). By observing the movement of S-stars orbiting around SgrA* and studying the behaviour of light near it, we can evaluate the accuracy of general relativity, our current theory of gravity, and test alternative theories that have not been observed before. This will further assist us in affirming or revising our knowledge of the nature of gravity in this region \cite{Gillessen2009}. 

In 2018, General Relativity (GR) was successfully confirmed through the detection of gravitational redshift near Sgr A*. This occurred upon the pericenter passage of one of the S-stars, namely S-2/S0-2 (the designations S-2 and S0-2 are intended for the European and American teams, respectively). The S0-2 star was at a semi-major axis of 1000au during its pericenter passage near Sgr A*\cite{Abuter2018}. Considering that the Schwarzschild radius/event horizon of Sgr A* is approximately 0.08au, a semi-major axis of 1000au still represents a considerable distance. However, to test the limits of GR, we would need to approach closer to the event horizon.  The presence of an extended distribution of mass, intermediate-mass black hole, and a nuclear star cluster within a few parsecs of the GC also affect the dynamics of the S-stars \cite{GGM2010, Do2020, Lalremruati2022b}. Several studies have used S-stars as a primary tool for theoretical and numerical simulations in order to understand the gravitational physics surrounding Sgr A* in greater detail \cite{Will2008, Merritt2010, Angelil2011, Zakharov2014, Lalremruati2021, Lalremruati2022a, Lalremruati2022b, Debo2023, PK2023}.

A binary star in a three-body interaction with a supermassive black hole results in the disintegration of the binary system where one of the stars is captured by the black hole and the other companion star is ejected from the vicinity of the black hole at a high velocity. Such stars ejected at a very high order velocity equal to or greater than V$\sim$ 1000$kms^{-1}$ are called hypervelocity stars (HVS). Meanwhile, the captured star has unusually high eccentricities and short orbital periods \cite{Hills1988}. Since 2005 several B-type hypervelocity stars (HVS) have also been detected near Sgr A*\cite{Brown2005}.  
Koposov et al. 2020 reported the fastest main sequence HVS  having a velocity of $\sim$ 1800$kms^{-1}$ when it was kicked out from the neighbourhood of Sgr A* \cite{Koposov2020}. The presence of HVS in close proximity to Sgr A* is an indication of a strong tidal interaction at the Galactic Center \cite{Sari2019,GGM2010}. Extreme tidal interactions can impact a star's structure and evolution by transferring energy, angular momentum, heating, mixing, or ejecting mass \cite{Alexander2003}. Perturbation in the orbit of stars due to the gravitational potential induces either a tidal disruption or tidal distortion of the stars depending upon their tidal radius \cite{Novikov1992,Rossi2021}.  A star that orbits within its tidal radius experiences tidal distortion while one that orbits at a distance lower than its tidal radius undergoes complete tidal disruption. The tidal distortion altogether affects the pericenter shift of the stellar orbit. Estimating the pericenter shift caused by the tidal distortion of S-stars near Sgr A* will provide additional precise information for any tests of gravitational theories near Sgr A* which use the pericenter shift of the S-stars as a probe. The pericenter shift due to the tidal distortion effect is  affected by (1) the pericenter distance ($r_p$) of the stellar orbit $\sim r_p^{-5}$(2) the mass-radius of the stars $\sim$ $M^{-1}R^5$ and (3) directly proportional to the tidal love number of the star, $k_2$ \cite{Will2008}. The tidal love number measures how much a self-gravitating object changes its shape in response to a gravitational potential. Augustus Edward Hough Love \cite{Love1909}  originally introduced the tidal love numbers in 1909 upon describing the ocean tides on Earth caused by its gravitational interaction with the Moon fully in the framework of Newtonian gravitational theory. Several factors such as mass, radius, elasticity and the equation of state determine the tidal love number of a stellar system \cite{Damour2009,Beuthe2015}. It has been reported that the tidal love number is heavily influenced by a star's polytropic index, with the tidal love number decreasing as the polytropic index increases \cite{Hinderer2008}. With all those consideration factors, we estimated the pericenter shift due to the tidal distortion effect of S-stars having highly eccentric orbit, e = 0.9. The high eccentricity is due to the reasons mentioned in \cite{Lalremruati2021}. The upper limit of the semi-major axis of the S-stars is chosen as 1000au, which is the semi-major axis of the S-02 star. Meanwhile, the lower bound for the semi-major axis is found to be 45au. This is obtained by considering the reported minimum age of the S-stars in the nuclear cluster near Sgr A* to be 6Myr \cite{Genzel2010} as well as the gravitational wave emission constraint given in \cite{MerrittGualandris2009} as
\begin{equation}
\begin{split}
a^4 = \frac{256G^3MM_{BH}(M+M_{BH})}{5c^5}(1-e^2)^{-7/2}\\
\Big(1+\frac{73}{24}e^2 + \frac{37}{96}e^4\Big)t_{GW}
\end{split}
\end{equation}
G, M, $M_{BH}$ denote the gravitational constant, mass of the star and mass of the black hole. In this present work, the mass-radius relationship for  Low-Mass Stars and High-Mass Stars is considered.  The tidal love number of Low-Mass Stars and High-Mass Stars  are obtained using a two-point Pade approximation for a certain range of polytropic indices.

 The paper is organized in the following way: Section II discusses the tidal love number formalism used in this work. Section III estimates the pericenter shift for several S-stars resulting from the tidal distortion effect. In Section IV, the estimated results are presented along with a brief discussion. Finally, Section V summarizes the findings of the study and outlines future prospects.
 
\section{Tidal love number }

A body, Z in perfect equilibrium under gravitational perturbation experiences a potential that includes both tidal, $U_{tidal}$  and centrifugal potentials, $\Phi_C$, given as  \cite{pw2014}
\begin{equation}
V = U_{tidal} + \Phi_C
\end{equation}
where,
\begin{equation}
U_{tidal} =  - \sum^{\infty}_{l=2} \frac{1}{l!} \xi_L(t)x^L 
\end{equation}
$\xi_L(t)$ is the symmetric-trace free (STF) tensor and equals the tidal moments associated with the external potential, $U_Z$ i.e \\

$\xi_L(t)$ :=  $- \delta_L U_{-Z}(t,0)$. \\

The tidal moments are taken to be a function of time and contain all the revelant information about the changing positions of the external bodies.
The centrifugal potential, $\Phi_C$ consists of monopole $(l = 0)$ and  quadrupole moment $(l = 2)$ i.e 
\begin{equation}
\Phi_C = \Phi^{l=0}_C + \Phi^{l=2}_C
\end{equation}
\\
 
However,  $l = 0$ corresponds to an unperturbed spherically symmetric system and the multipole expansion starts from $l = 2$. Thus, the monopole contribution to the centrifugal potential is ignored here, i. e
\begin{equation}
\Phi_C = \Phi^{l=2}_C = \frac{1}{6}w^2 (x^2 +y^2 - 2z^2) = -\frac{1}{3}w^2r^2P_2(cos\theta)
\end{equation}

\begin{figure}[b]
 \includegraphics[width=0.4\textwidth]{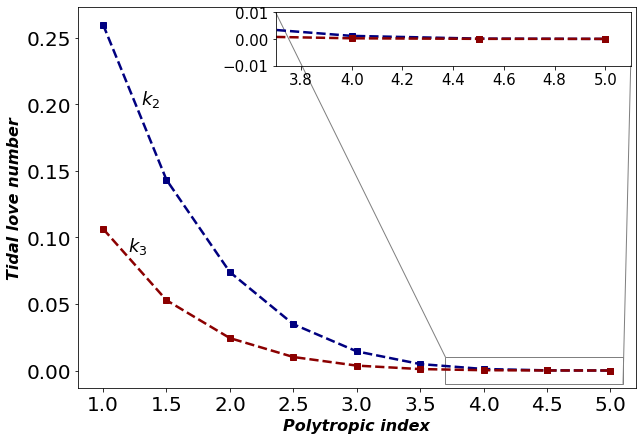}
 \caption{\label{fig:klvrsn}Tidal love number $(k_l)$ plotted against polytropic index of stars for multipole moment $l = 2$ and $ l =3$.}
 \end{figure}

$'w'$ is the angular-velocity vector and $P_2(cos\theta)$is the Legendre Polynomial. 
 Considering the above conditions, equation(2) can be re-written as
 
 \begin{equation}
 V =  \sum_{lm} \frac{4\pi}{2l+1}d_{lm}r^lY_{lm}(\theta, \phi)
 \end{equation}
 
 where, $Y_{lm}(\theta, \phi)$ denotes the spherical harmonics. The total gravitational potential of the deformed body can then be interpreted as

\begin{equation}
U = -\frac{GM}{r} + \sum_{lm} \frac{4\pi}{2l+1} \frac{2k_lR^{2l+1}d_{lm}}{r^{l+1}}Y_{lm}(\theta,\phi)
\end{equation}

$'k_l'$ denotes the tidal love number which depends on the details of the unperturbed configuration. The tidal love number is also deeply affected by the polytropic index of stars. 
The polytropic equation gives the simplest stellar model and is given by the famous Lane-Emden model where the pressure(P) and density$(\rho)$ are related as  \cite{Paddy}
\begin{equation}
\rho = KP^{\gamma}
\end{equation}
 where, $\gamma$ = $(1+\frac{1}{n})$, 'n' here is the polytropic index of a star. The value of 'n' represents how rapidly pressure and density fall from the center to the surface of the star and ultimately determines the stability of the star. Different values of 'n' represents different type of stars.  This work considers a polytropic index, n = 1.0, 1.5, 2.0, 2.5, 3.0, 3.5, 4.0, 4.5 and 5.0.
 \begin{figure}[b]
 \includegraphics[width=0.4\textwidth]{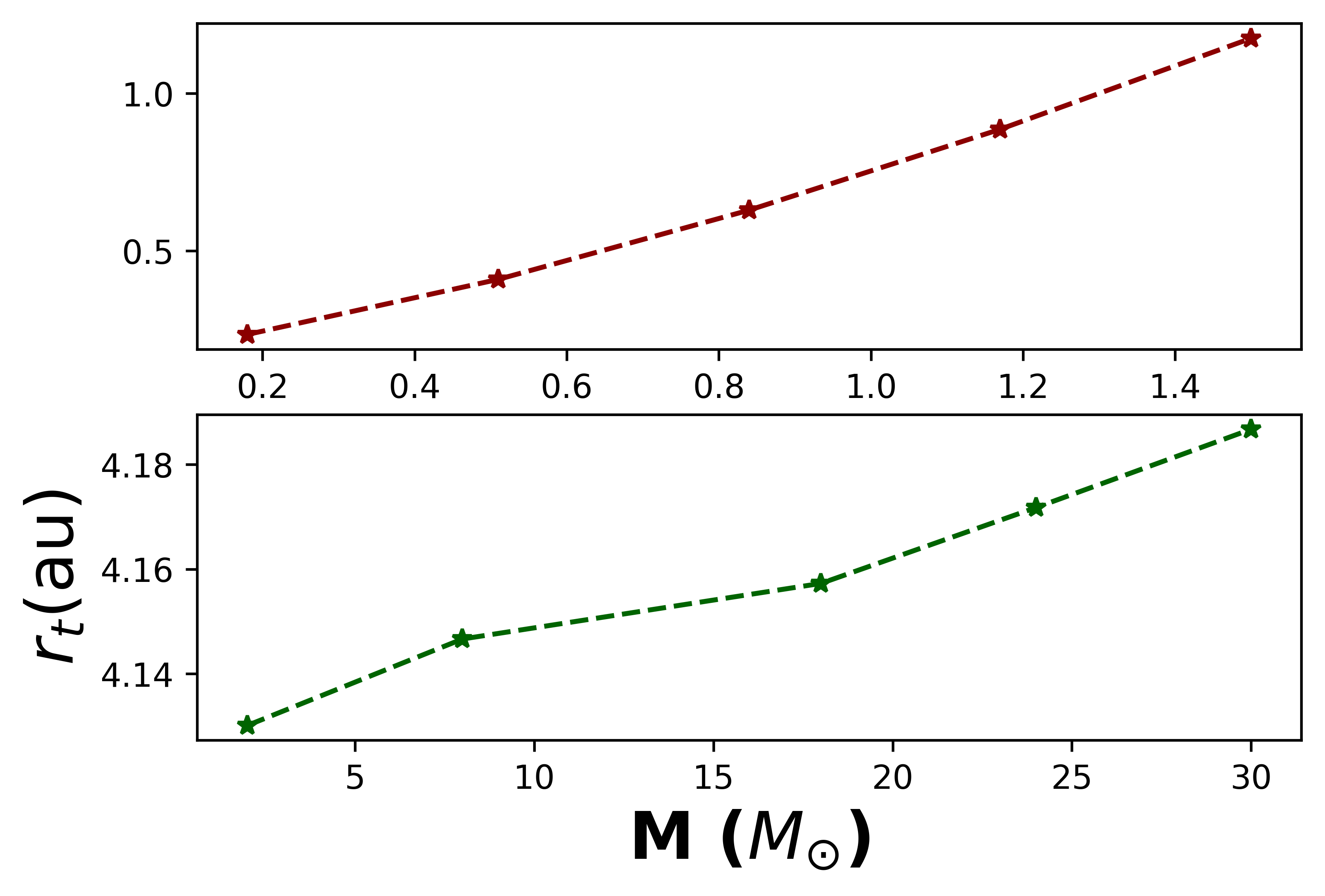}
 \caption{\label{fig:tidalradii} Tidal radii plotted against its respective stellar masses. The upper plot is for Low-Mass Stars (0.18$M_{\odot}$ $\le$ 1.5$M_{\odot}$) and the lower plot is for High-Mass Stars (1.5$M_{\odot}$ $\le$ 30$M_{\odot}$) .}
 \end{figure}  
 
 To get a sense of how the polytropic index affects  the stability of a star, Yip and Leung using a perturbative expansion gave a  Pade approximation of the Love number  \citep{YipLeung2017}
 
 \begin{equation}
 k_l(n) = (5-n)^3\frac{a_1 + na_2 +n^2a_3 +n^3a_4}{1000+na_5+n^2a_6}
 \end{equation}
 $a_1$, $a_2$, .... $a_6$ are coefficients determined from the perturbation series and depending upon their multipole moments as displayed in Table 1 of \cite{YipLeung2017}. This work considers multipole moments, $l = 2$ and $l = 3$.
 
\section{Pericenter shift due to tidal effect}

S-stars in an elliptical orbit around the supermassive black hole Sgr A* experience a  precession. The pericenter point of the S-stars rotates along the motion of Sgr A*. The pericenter shift of the S-stars can be used as a probe for astronomical tests of gravitational theories such as general relativity, $f(R)$ gravity, Kaluza- Klein gravity ... etc near Sgr A*\citep{ Lalremruati2021, Lalremruati2022a, Lalremruati2022b, Debo2023}. Each gravitational theory has several factors contributing to the pericenter shift of the S-stars. However, in this work, we are interested only in pericenter shift due to tidal distortion of S-stars near Sgr A*. A star with a highly eccentric and compact orbit around a black hole surviving a complete tidal disruption may undergo tidal distortion due to the black hole's gravitational pull. The tidal radii, $r_t$, determining the fate of the stars is expressed as \citep{Hills1975,Rossi2021,Alex2005}

\begin{equation}
r_t = (\frac{M_{BH}}{M})^{1/3}R
\end{equation}
 
 '$M_{BH}$' here is the mass of the black hole, while 'M' and 'R' are the mass and radius of the stars. The tidal radii vary accordingly with stellar mass as shown in Fig. 2.

  The tidal distortion effect contribution arises when the considered  S-stars are in highly eccentric orbits near Sgr A*. The estimated tidal distortion effect can also be added to pericenter shift contribution of several gravitational theories. This indicates the prime importance of estimating the pericenter shift due to the tidal distortion effect. The corresponding pericenter shift due to tidal distortion is given by \citep{Will2008}
 
\begin{figure}[b]
 \includegraphics[width=0.5\textwidth]{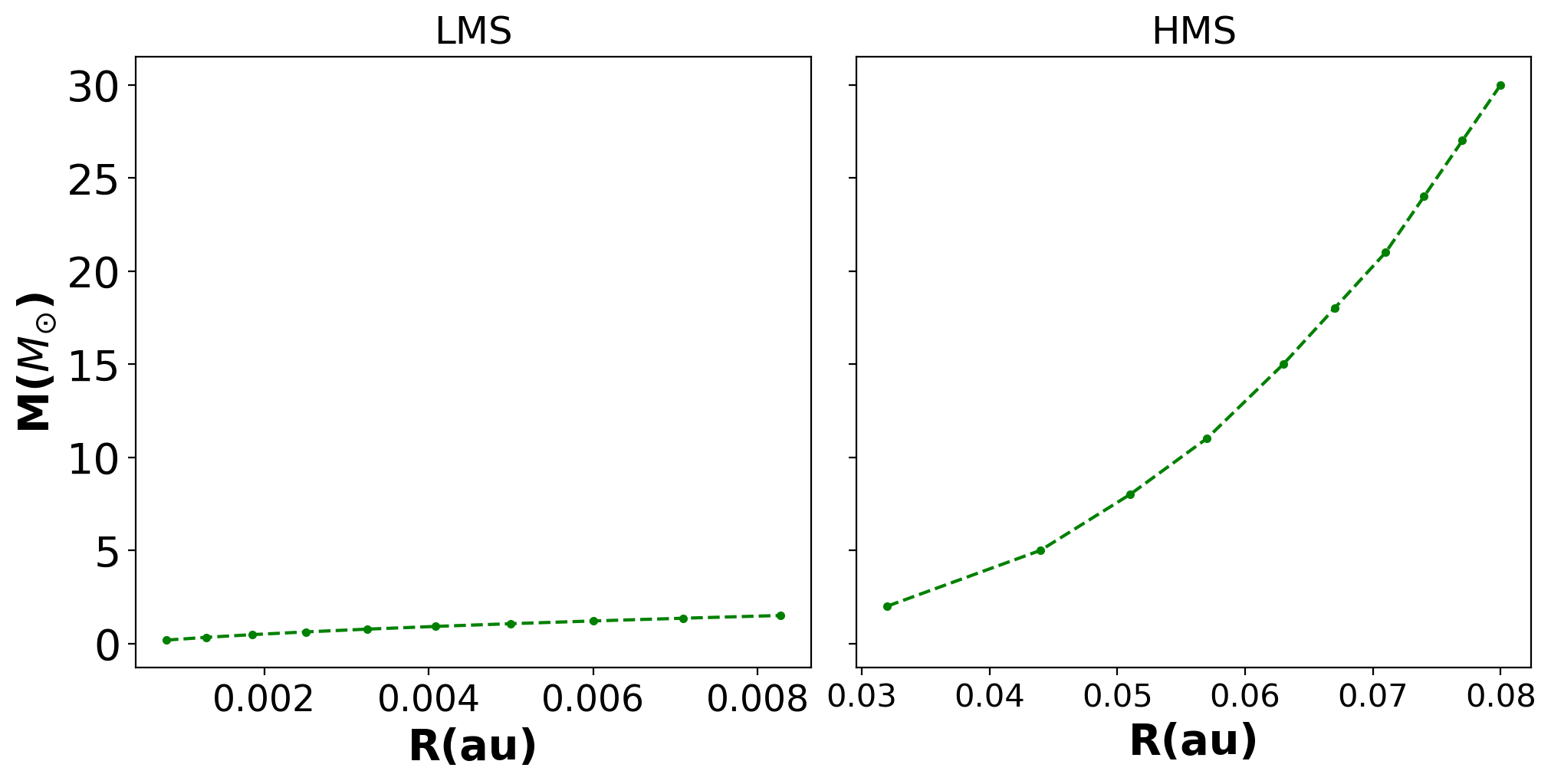}
 \caption{\label{fig:MR} Mass-Radius plot for Low Mass Stars (LMS) and High Mass Stars (HMS).}
 \end{figure}

 \begin{equation}
\delta\theta^{tidal}_{prec} = \frac{30\pi}{(1+e)^5}k_2(1+\frac{3e^2}{2}+\frac{e^4}{8})(\frac{M_{BH}}{M})(\frac{R^5}{r_p^5})
\end{equation}
    
    \begin{figure*}
     \centering
     \begin{subfigure}{0.42\linewidth}
         \includegraphics[width=\linewidth]{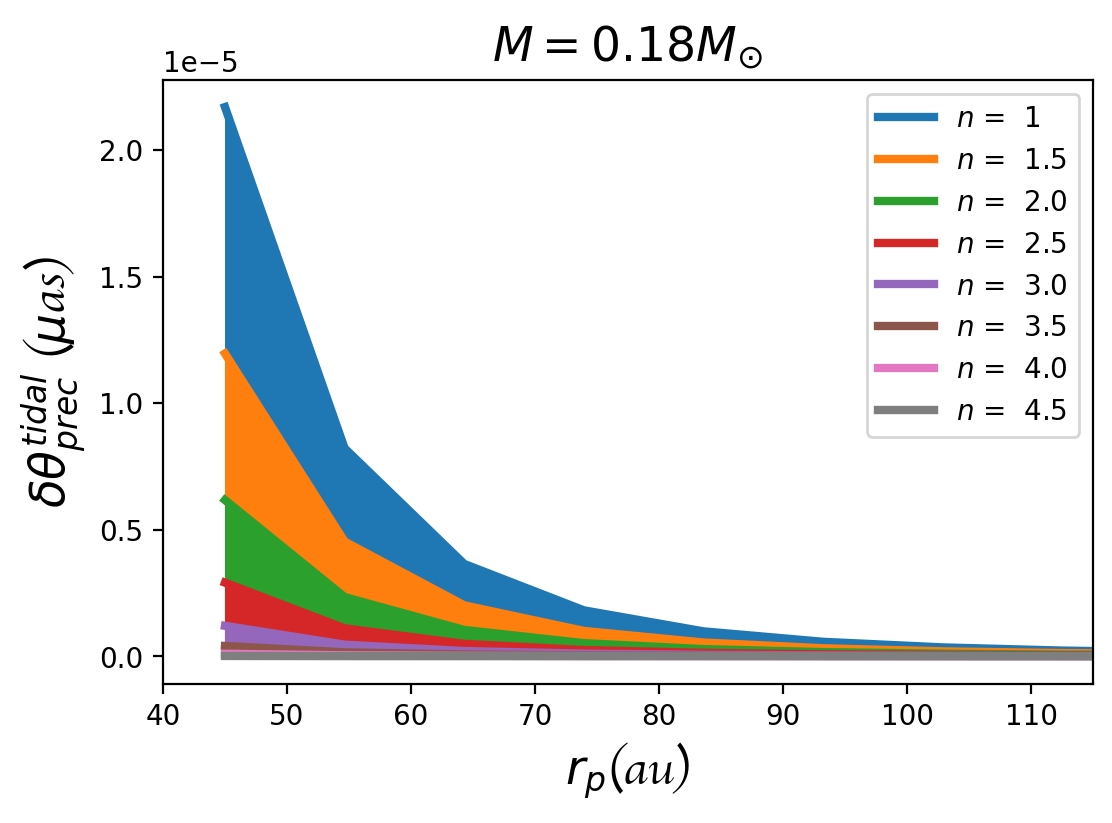}
             \end{subfigure} \hfill
         \begin{subfigure}{0.45\linewidth}
         \includegraphics[width=\linewidth]{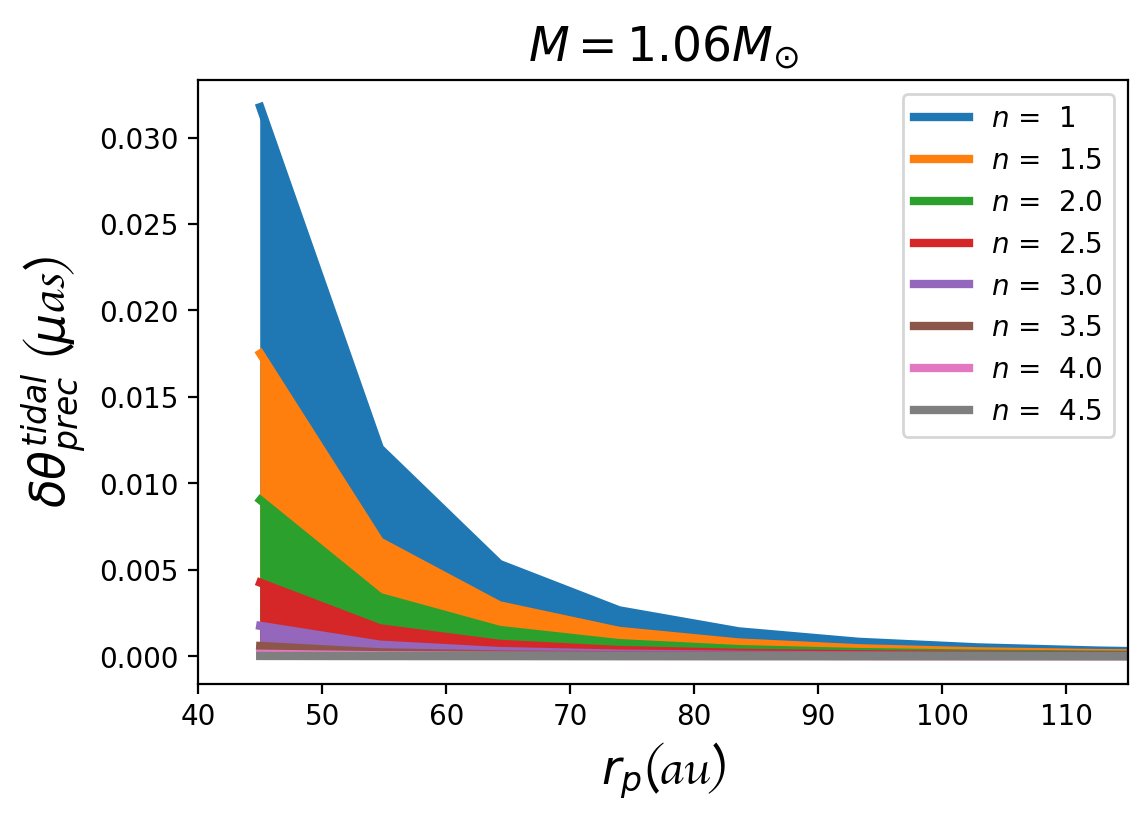}
             \end{subfigure}\hfill
            \begin{subfigure}{0.45\linewidth}
         \includegraphics[width=\linewidth]{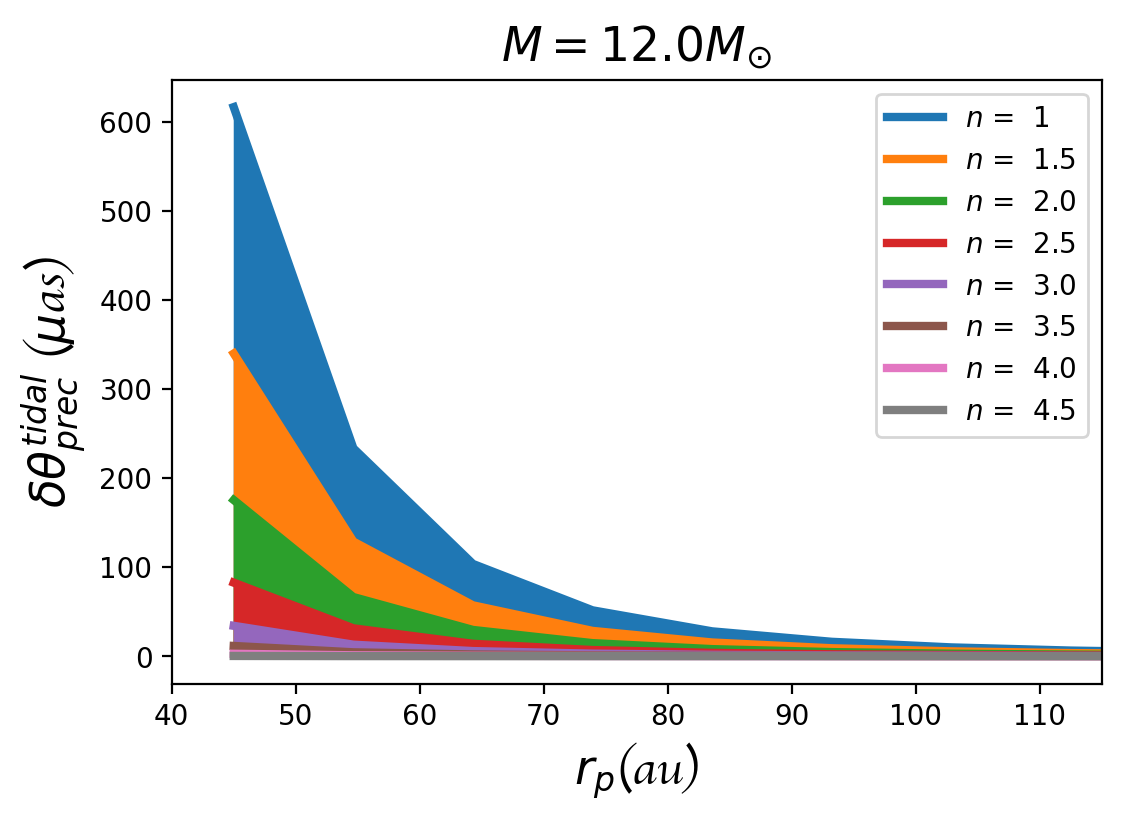}
                           \end{subfigure}\hfill
         \begin{subfigure}{0.45\linewidth}
         \includegraphics[width=\linewidth]{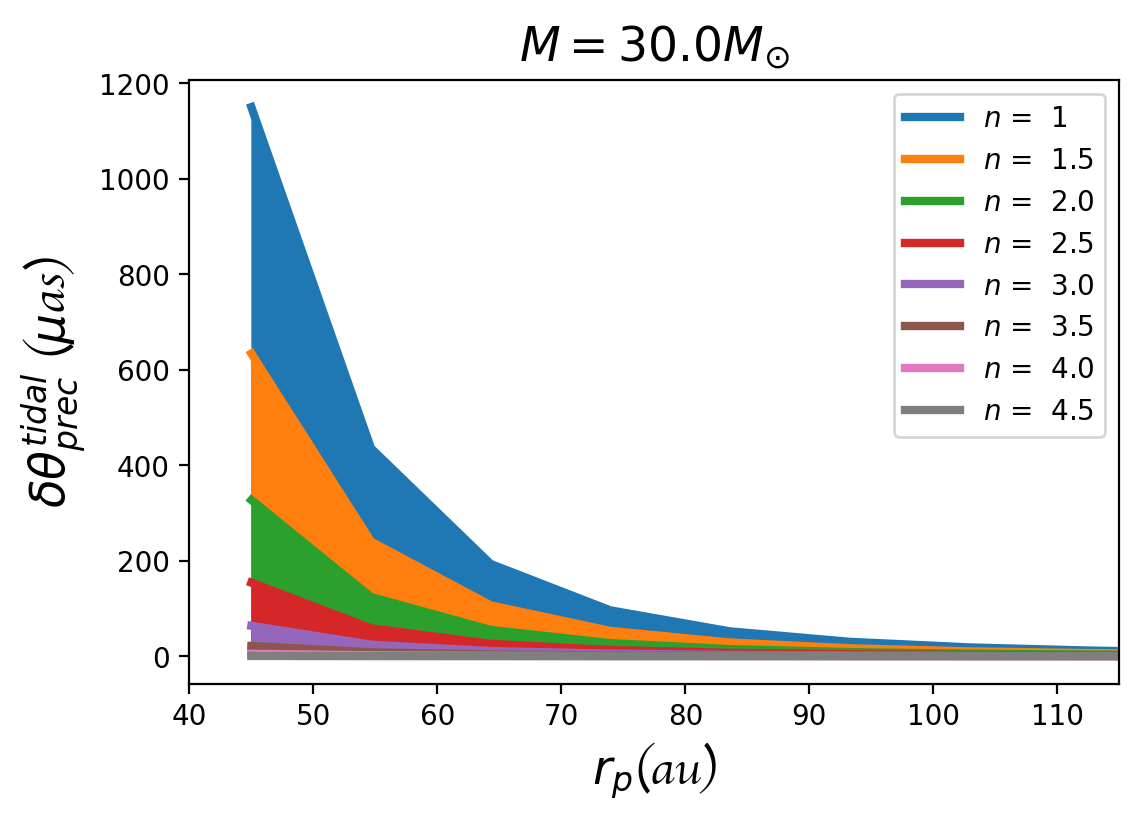}
          \end{subfigure}
              \caption{Pericenter shift due to tidal distortion effect plotted against the pericenter distance, $r_p$ for Low Mass Stars (LMS)(0.18$M_{\odot}$ \& 1.06$M_{\odot}$) and High Mass Stars (HMS) (12$M_{\odot}$ \& 30.0$M_{\odot}$ ) at $k_2$. The polytropic indices, n = 1, 1.5, 2.0, 2.5,  3.0, 3.5, 4.0 and 4.5 are considered for each stellar mass.}
        \label{fig:}
\end{figure*}

\begin{figure*}
     \centering
     \begin{subfigure}{0.42\linewidth}
         \includegraphics[width=\linewidth]{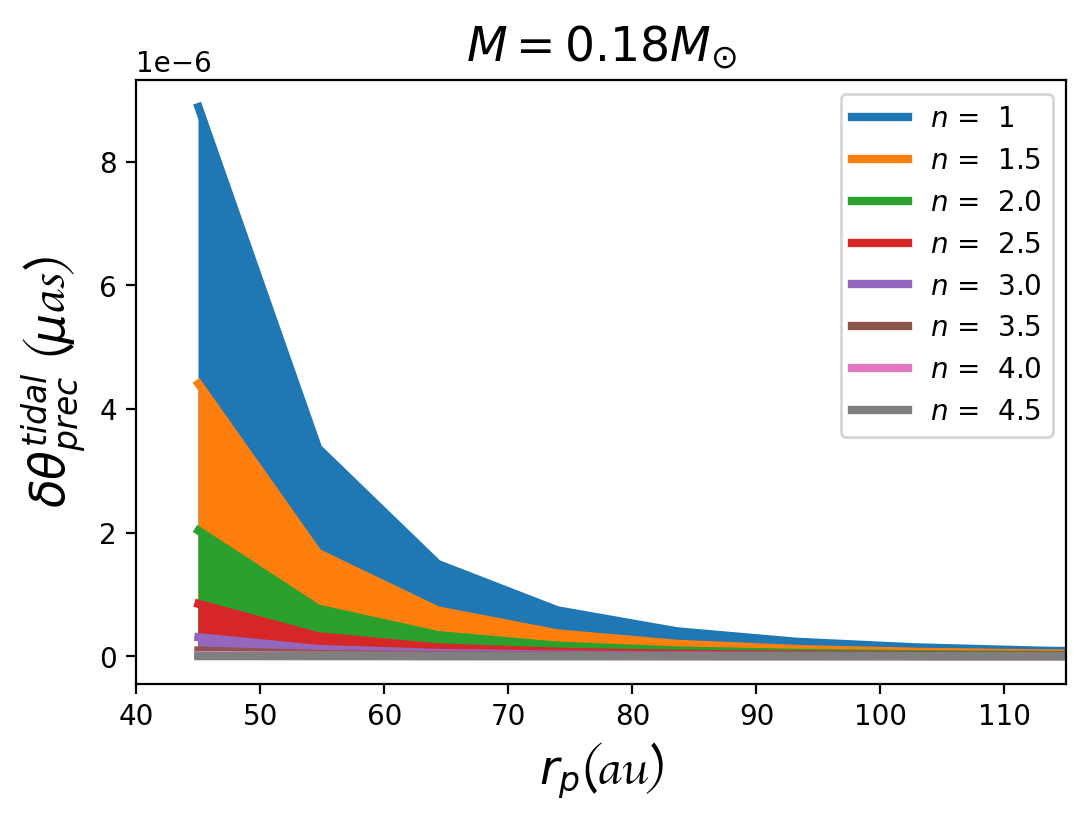}
           \end{subfigure} \hfill  
          \begin{subfigure}{0.45\linewidth}
         \includegraphics[width=\linewidth]{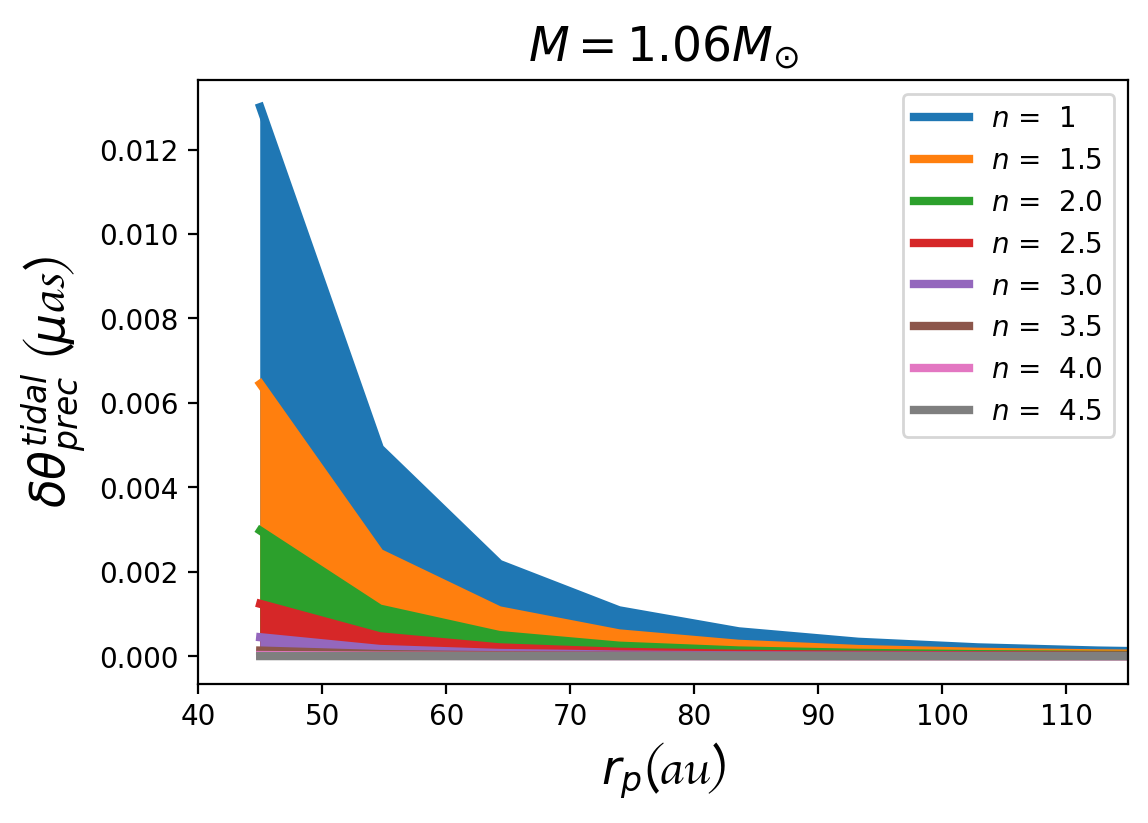}           
     \end{subfigure}\hfill     
     \begin{subfigure}{0.45\linewidth}
         \includegraphics[width=\linewidth]{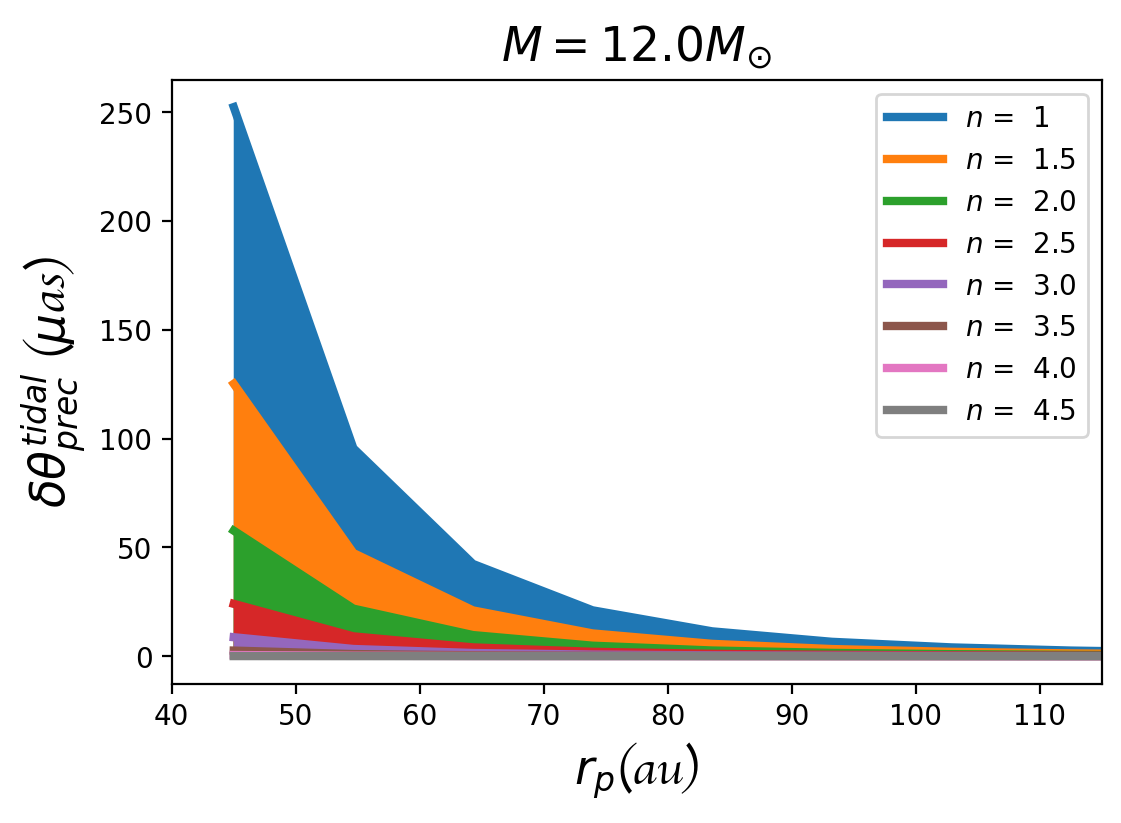}
                           \end{subfigure}\hfill
               \begin{subfigure}{0.45\linewidth}
         \includegraphics[width=\linewidth]{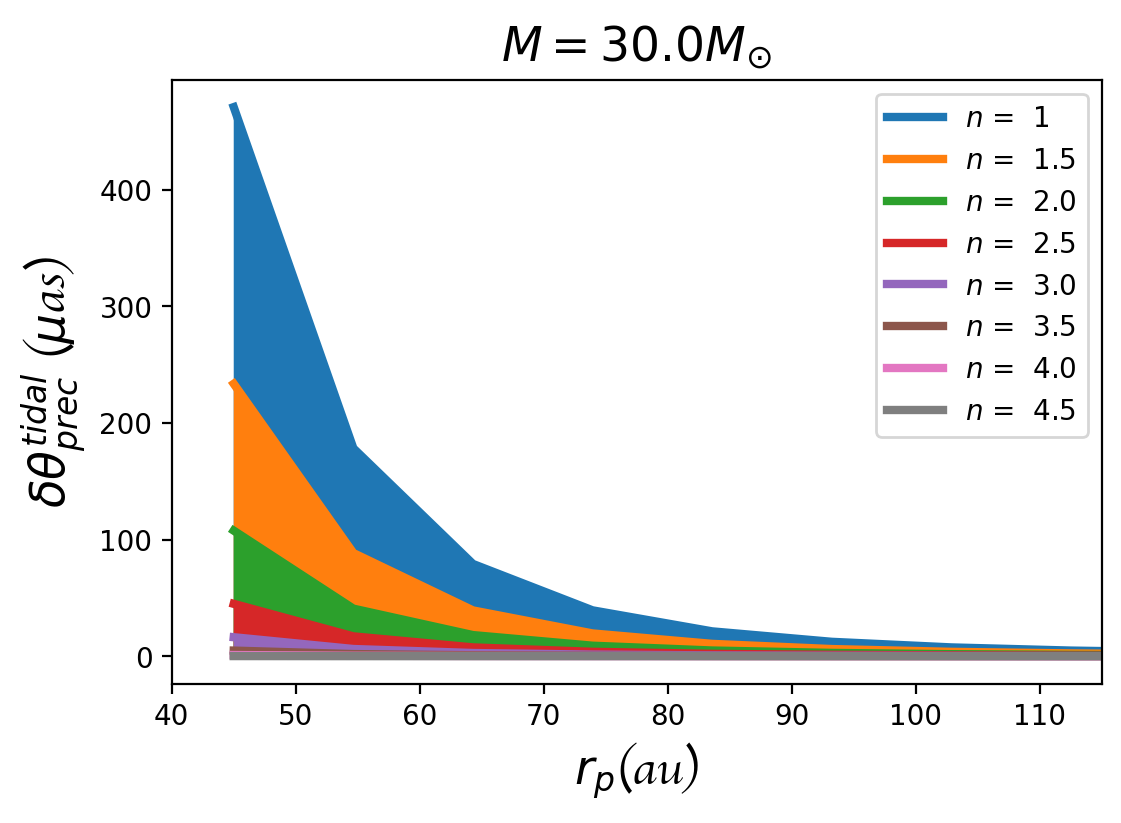}
                   \end{subfigure}
              \caption{Pericenter shift due to tidal distortion effect plotted against the pericenter distance, $r_p$ for Low Mass Stars (LMS)(0.18$M_{\odot}$ \& 1.06$M_{\odot}$) and High Mass Stars (HMS) ( 12$M_{\odot}$ \& 30.0$M_{\odot}$ ) at $k_3$. The polytropic indices, n = 1, 1.5, 2.0, 2.5,  3.0, 3.5, 4.0 and 4.5 are considered for each stellar mass.}
        \label{fig:}
\end{figure*}

\begin{figure*}
     \centering
     
         \includegraphics[width=\linewidth]{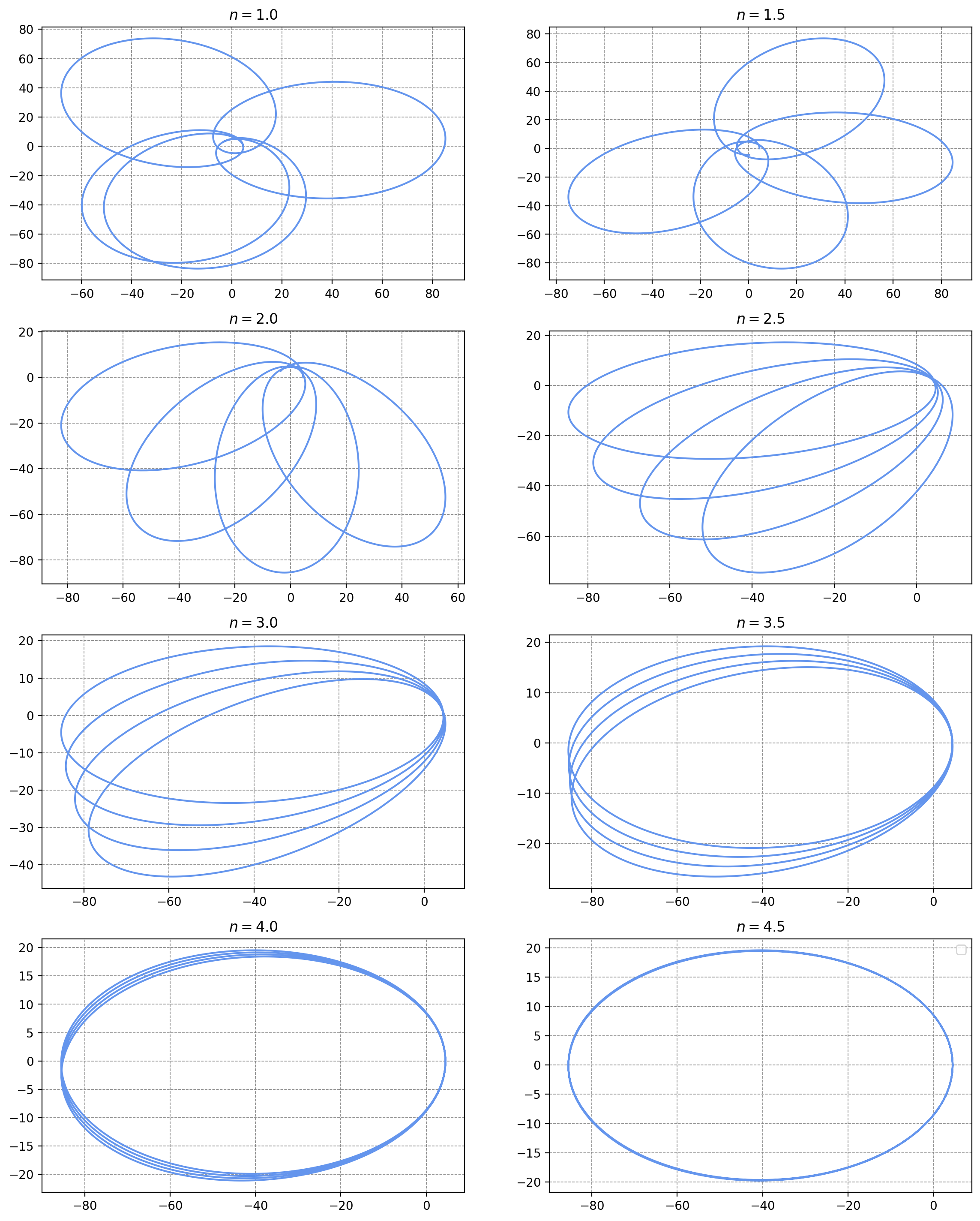}
                     \caption{Precession due to tidal distortion effect of S-stars of mass, M = 0.18$M_{\odot}$  near Sgr A* having different polytropic indices, n = 1.0, 1.5, 2.0, 2.5, 3.0, 3.5, 4.0 and 4.5 at a semi-major axis of 45au ($r_p$ = 4.5au). }
        \label{fig:}
\end{figure*}    

The tidal effect is highly governed by the eccentricity, love number and mass-radius relationship of the S-stars. In this work, we consider the Mass-Radius Relation(MRR) for Low-Mass Stars (LMS) and High-Mass Stars (HMS). LMS have masses in the range 0.18 $\le$ $M/M_{\odot}$ $\le$ 1.5 and the HMS ranges between 1.5 $\le$ $M/M_{\odot}$ $\le$ 30. 
Eker and his team in 2018 used data from 509 main-sequence stars in the solar neighbourhood and obtained the Mass-Luminosity Relation(MLR), Mass effective Temperature Relation(MTR) and MRR for LMS. The MRR for LMS(0.18 $\le$ $M/M_{\odot}$ $\le$ 1.5) is given as \citep{Eker2018} 
\begin{equation}
\frac{R}{R_{\odot}} \cong [0.438 (\frac{M}{M_{\odot}})^2 + 0.479 (\frac{M}{M_{\odot}})+ 0.075]
\end{equation}


\begin{table*}
\caption{\label{tab:table3}Estimated rate of pericenter shift ($\dot\theta^{tidal}_{prec}$) at $\mu$as/yr for polytropic indices, n = 1, 1.5, 2.0, 2.5, 3.0, 3.5, 4.0 and 4.5 for LMS and HMS at pericenter distance of 4.5 au. M denotes the stellar masses while $k_2$ and $k_3$ are the tidal love number at multipole moments $l=2$ and $l =3$. }
\begin{ruledtabular}
\begin{tabular}{ccccccccccccc}
&\multicolumn{6}{c}\textbf{\large{$\dot\theta^{tidal}_{prec}$ ($\mu$as/yr)} } \\
\\
 \textbf{$polytropic$}&\multicolumn{6}{c}{$k_2$}&\multicolumn{4}{c}{$k_3$}\\
 {$index$}&&\multicolumn{9}{c}M\\
 \\
 \\
 $n$&0.18$M_{\odot}$&0.9$M_{\odot}$&1.5$M_{\odot}$&7.8$M_{\odot}$&17.3$M_{\odot}$&30$M_{\odot}$&0.18$M_{\odot}$&0.9$M_{\odot}$&1.5$M_{\odot}$&7.8$M_{\odot}$&17.3$M_{\odot}$&30$M_{\odot}$  \\ \hline
 \\
 \\
 1&4e-08&2.3e-05&5.2e-05&0.86&1.48&2.15&1.6e-08&9.4e-06&2.1e-0.4&0.35&0.61&0.88      \\
 \\
 \\
 1.5&2.2e-08&1.3e-05&2.9e-04&0.47&0.81&1.18&8.1e-09&4.7e-06&1.1e-04&0.17&0.3&0.44\\
 \\
 \\
 2.0&1.1e-08&6.6e-06&1.5e-04&0.24&0.42&0.61&3.7e-09&2.2e-06&4.9e-05&0.08&0.14&0.2\\
 \\
 \\
 2.5&5.3e-09&3.1e-06&7.0e-05&0.11&0.20&0.29&1.6e-09&9.0e-07&2.03e-05&0.33&0.058&0.08\\
 \\
 \\
 3.0&2.2e-09&1.3e-06&2.9e-05&0.047&0.082&0.12&5.6e-10&3.3e-07&7.4e-06&0.012&0.021&0.03\\
 \\
 \\
 3.5&7.4e-10&4.3e-07&9.7e-06&0.016&0.028&0.04&1.7e-10&9.6e-08&2.2e-06&0.0036&0.006&0.009\\
 \\
 \\
 4.0&1.7e-10&1.0e-07&2.3e-06&0.0037&0.0064&0.0094&3.4e-11&2.0e-08&4.5e-07&7.0e-04&0.0013&0.002\\
 \\
 \\
 4.5&1.7e-11&9.8e-09&2.2e-07&3.7e-04&6.3e-04&9.1e-04&3.14e-12&1.8e-09&4.1e-08&6.8e-05&1.2e-04&0.0002\\
 
 \\
 \\
 
\end{tabular}
\end{ruledtabular}
\end{table*}

For HMS only the MTR and MLR were obtained in \citep{Eker2018} . Later in 2021, using the MTR and MLR for HMS obtained by \citep{Eker2018}, Lalremruati \& Kalita obtained the MRR for HMS (1.5 $\le$ $M/M_{\odot}$ $\le$ 30) given as \cite{Lalremruati2021}

\begin{equation}
log(\frac{R}{R_{\odot}}) \cong [-0.3435log(\frac{M}{M_{\odot}})+ 0.34 log(\frac{M}{M_{\odot}})^2+ 0.7365]
\end{equation}

Fig. 3. shows a plot for the M-R relationship for LMS and HMS. It can be seen that for LMS the stellar mass varies linearly with its radius. However, for HMS, the stellar mass varies exponentially with its radius.

 For estimating the pericenter shift due to tidal distortion effect, LMS of 0.18 $M_{\odot}$, 0.62  $M_{\odot}$, 1.06  $M_{\odot}$ and 1.5  $M_{\odot}$  and HMS of 3  $M_{\odot}$, 12  $M_{\odot}$, 21  $M_{\odot}$ and 30  $M_{\odot}$ are considered. 
 
  The astrometric size of the rate of pericenter shift due to tidal distortion at the pericenter, $r_p$ is expressed as
  \begin{equation}
  \dot\theta^{tidal}_{prec} = \delta\theta^{tidal}_{prec}(\frac{r_p Sin(i)}{PD})
   \end{equation}
   where,
   
   $i$ is the inclination angle and is taken to be $90^\circ$ to get the maximum contribution, 
   
    $P = \frac{2\pi r_p^{3/2}}{\sqrt{GM_{BH}}(1-e)^{3/2}}$ is the period of the stars and 
    
    $D$ is the Earth - Galactic Center distance

\section{Results and discussion}

The tidal love number affecting the pericenter shift of S-stars is highly influenced by the polytropic index of the stellar system as shown in Figure 1. The figure displayed a plot of tidal love number $k_2$ (for multipole moment $l = 2$) and $k_3$(for multipole moment $l = 3$) against their polytropic indices. We see that at a lower polytropic index, $k_2$ dominates over $k_3$. But, as the polytropic indices reach $\sim$ 3.7, the tidal love number is independent of its multipole moment.

For tidal love number with multipole moment $l = 2$ i.e. $k_2$, we can see from Figure 4. that the pericenter shift due to the tidal distortion effect gradually increases as we increase the stellar mass.  We can also see that the pericenter shift varies with the polytropic indices and increases with decreasing polytropic indices. However, beyond a pericenter distance of $r_p$ $\sim$ 110au, the estimated pericenter shift due to tidal distortion is independent of its polytropic indices.  The above obtained results are more or less the same for the tidal love number with multipole moment $l = 3$ i.e. $k_3$ which can be seen in Figure 5.
 
From Table I, we can see that at a pericenter distance, $r_p$ = 4.5 au (which is the minimum possible distance for a star to have a stable orbit around the GC supermassive black hole, Sgr A* ) the rate of pericenter shift due to tidal distortion effect is highest at around   M = 30$M_{\odot}$ and n = 1. Polytropes n = 1.5 \& n =2.0  also shows promising results. This shows that the rate of the pericenter shift increases with increasing stellar mass and decreasing polytropic indices. The results are the same for $k_2$ and $k_3$. However,   we can see from Table$I$  that $k_2$ obtained a higher rate of pericenter shift as compared to $k_3$. This is also true upon estimating the pericenter shift as shown in Figure. 4 and Figure 5. Visualization of the precession due to the tidal distortion effect at a semi-major axis of 45 au near Sgr A* is displayed in Figure 6 for a stellar mass of M = 0.18$M_{\odot}$. The precession is displayed for different stellar polytropic indices, n = 1.0, 1.5, 2.0, 2.5, 3.0, 3.5, 4.0 and 4.5. The precession becomes more prominent as the polytropic index decreases for stars.

\section{Conclusions and future prospects}

To obtain the maximum pericenter shift due to the tidal distortion effect well below  $r_p $ $\sim$ 100au (a $\sim$ 1000au) we need HMS to have lower polytropic indices possibly in the ranges, of n = 1.0 - 2.5.  The multipole moments of the tidal love number also play a crucial role in estimating the pericenter shift due to the tidal distortion effect. The tidal love number, $k_2$ (having multipole moment $l = 2$) yields better and higher results than $k_3$ with multipole moment $l =3$ as seen in Fig. 4, Fig. 5 and Table $I$. The pericenter shift due to the tidal distortion effect is higher at compact orbits and eventually decreases in wider orbits. The estimated results provided an independent test of gravity near Sgr A*. The results can then be added to any tests of gravitational theories near Sgr A* using the pericenter shift as a main probe. The astrometric capabilities of the existing telescopes such as Keck, Very Large Telescope (VLT) and the upcoming Extremely Large Telescope (ELT) have the capabilities to constrain the tidal distortion effect of S-stars lying at a compact orbit of Sgr A*. Thus, this will eventually contribute to the understanding of the nature of gravity in such extreme regions of gravity such as the immediate neighbourhood of a supermassive black hole.

\section{Acknowledgement}

The research of Zodinmawia is funded by DST-SERB project Sanction No. EEQ/2021/00385.

\bibliographystyle{apsrev4-2}
\bibliography{aurora}

\end{document}